\documentclass[sigconf, nonacm]{acmart}

\newcommand\vldbdoi{10.14778/3632093.3632115}
\newcommand\vldbpages{553 - 562}
\newcommand\vldbvolume{17}
\newcommand\vldbissue{3}
\newcommand\vldbyear{2023}
\newcommand\vldbauthors{\authors}
\newcommand\vldbtitle{\shorttitle} 
\newcommand\vldbavailabilityurl{URL_TO_YOUR_ARTIFACTS}
\newcommand\vldbpagestyle{empty} 

\usepackage{amsmath}
\DeclareMathOperator*{\argmax}{arg\,max}
\usepackage{svg}
\usepackage{multirow}
\usepackage{enumitem}
\usepackage{adjustbox}
\usepackage{caption}
\usepackage{subcaption}

\newcommand{\tp}[1]{{\color{red} {\bf ??? #1 ???}}\normalcolor}

\begin{document}
\title{ADF \& TransApp: A Transformer-Based Framework for Appliance Detection Using Smart Meter Consumption Series}

\author{Adrien Petralia}
\affiliation{%
  \institution{EDF R\&D - Université Paris Cité}
}
\email{adrien.petralia@gmail.com}

\author{Philippe Charpentier}
\affiliation{%
  \institution{EDF R\&D}
}
\email{philippe.charpentier@edf.fr}

\author{Themis Palpanas}
\affiliation{%
  \institution{Université Paris Cité - IUF}
}
\email{themis@mi.parisdescartes.fr}
\begin{abstract}
Over the past decade, millions of smart meters have been installed by electricity suppliers worldwide, allowing them to collect a large amount of electricity consumption data, albeit sampled at a low frequency (one point every 30min). 
One of the important challenges these suppliers face is how to utilize these data to detect the presence/absence of different appliances in the customers' households. 
This valuable information can help them provide personalized offers and recommendations to help customers towards the energy transition. 
Appliance detection can be cast as a time series classification problem. 
However, the large amount of data combined with the long and variable length of the consumption series 
pose challenges when training a classifier.
In this paper, we propose ADF, a framework that uses subsequences of a client consumption series to detect the presence/absence of appliances. 
We also introduce TransApp, a Transformer-based time series classifier that is first pretrained in a self-supervised way to enhance its performance on appliance detection tasks. 
We test our approach on two real datasets, including a publicly available one.
The experimental results with two large real datasets show that the proposed approach outperforms current solutions, including state-of-the-art time series classifiers applied to appliance detection.
This paper appeared in VLDB 2024.
\end{abstract}

\maketitle

\pagestyle{\vldbpagestyle}
\begingroup\small\noindent\raggedright\textbf{PVLDB Reference Format:}\\
\vldbauthors. \vldbtitle. PVLDB, \vldbvolume(\vldbissue): \vldbpages, \vldbyear.\\
\href{https://doi.org/\vldbdoi}{doi:\vldbdoi}
\endgroup
\begingroup
\renewcommand\thefootnote{}\footnote{\noindent
This work is licensed under the Creative Commons BY-NC-ND 4.0 International License. Visit \url{https://creativecommons.org/licenses/by-nc-nd/4.0/} to view a copy of this license. For any use beyond those covered by this license, obtain permission by emailing \href{mailto:info@vldb.org}{info@vldb.org}. Copyright is held by the owner/author(s). Publication rights licensed to the VLDB Endowment. \\
\raggedright Proceedings of the VLDB Endowment, Vol. \vldbvolume, No. \vldbissue\ %
ISSN 2150-8097. \\
\href{https://doi.org/\vldbdoi}{doi:\vldbdoi} \\
}\addtocounter{footnote}{-1}\endgroup

\ifdefempty{\vldbavailabilityurl}{}{
\vspace{.3cm}
\begingroup\small\noindent\raggedright\textbf{PVLDB Artifact Availability:}\\
The source code, data, and/or other artifacts have been made available at \url{https://github.com/adrienpetralia/TransApp}. 
\endgroup
}

\section{Introduction}
\label{sec:intro}

Over the past ten years, smart meters have been installed in millions of households across the globe by electricity providers~\cite{smart_meter_deployments_ue, smart_meter_deployments_us}. 
These meters capture detailed time-stamped electricity consumption data at a very low frequency (for instance, 15min in Italy, 30min in the UK, 30min in France and 60min in Spain~\cite{ZHAO2020114949}) that enable consumers to understand better and manage their electrical consumption~\cite{smartmeterinsmartgrid}.


\begin{figure}
    \centering
    \includegraphics[width=1\linewidth]{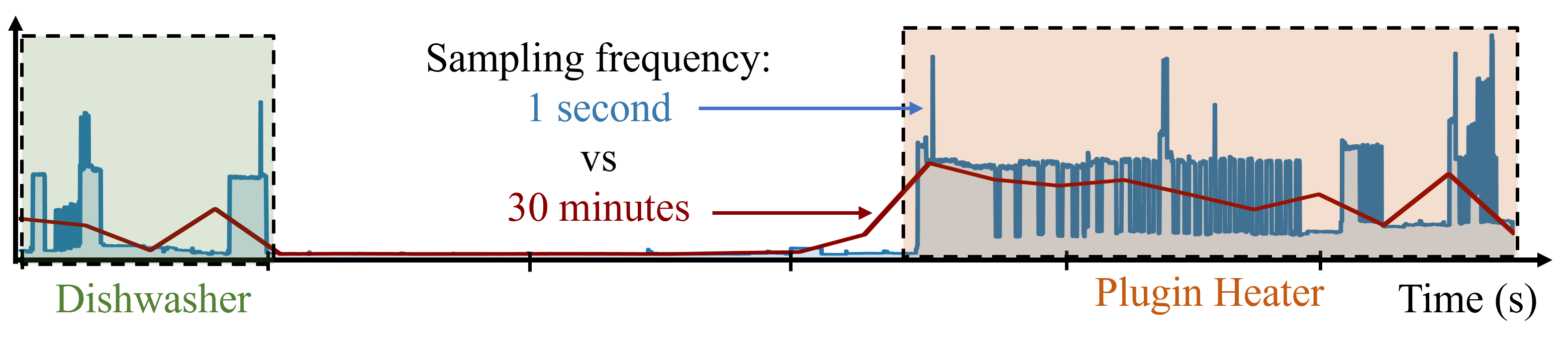}
     \vspace*{-0.8cm}
    \caption{Example of a consumption series of 12hours, containing a dishwasher and a plugin heater at two sampling frequencies (1 second vs. 30min).}
     \vspace*{-0.8cm}
    \label{fig:introfig} 
\end{figure}

For electricity suppliers, it is crucial to know whether their customers own specific electrical appliances. 
This knowledge allows them to segment customers~\cite{ALLCOTT20111082}, offering personalized services that increase their satisfaction and retention and help customers rationalize their electricity consumption, contributing to the energy transition. 
One approach to gather this information is through a consumption questionnaire, which can be time-consuming, costly, and prone to errors. 
To overcome these challenges, electricity suppliers can use advanced data analytics techniques to detect appliances directly from the smart meter data collected. This approach is more efficient and less intrusive, enabling suppliers to gather the necessary information without inconveniencing their customers.

Appliance detection has become an important area of research~\cite{LIU2021283, ROSSIER2017691, dengetal, appliancedetectionbenchmark}. 
Signature-based methods are widely adopted and use information related to the unique patterns of specific appliances~\cite{huquet_activation}.
However, most of these studies relied on data from smart meters capable of recording one, or more values per second, in contrast to the vast majority of smart meters installed by suppliers that collect data at considerably lower frequencies. 
As shown in Figure~\ref{fig:introfig}, low sampling rates result in a smoothed signal that does not maintain the unique pattern information of each appliance. 
One way to tackle this challenge is to cast the appliance detection problem as a binary classification task, where a time series classifier is trained to detect an appliance in a consumption series.
Few recent studies have tried this approach using very low-frequency smart meter data (one point sampled every minute or less) and show promising results~\cite{Albert_2013, dengetal, appliancedetectionbenchmark}.
Nevertheless, the series collected by suppliers are really long (10k-20k points), in large numbers (i.e., from many customers), and of variable length: not all classifiers can handle data with such characteristics, and training a model with them can be computationally intensive. 
One way to handle this issue is to fragment the total consumption series in subsequences of equal length.
This process has been investigated in~\cite{dengetal} and leads to superior results than using the full consumption series as input.
In our approach, we use the same strategy.


In recent years, the Transformer-based architecture 
has been investigated for time series, with promising results on downstream tasks~\cite {tst_survey, zerveas_tst}.
One of the advantages of these architectures is their ability to be first trained on an auxiliary task to learn a representation, and then finetuned to perform better on downstream tasks with the same amount of labeled data~\cite{bert, zerveas_tst, Yuqietal-2023-PatchTST}.
Given that electricity suppliers collect large amounts of non-labeled data from their clients, this type of architecture and two step training process are good candidates to enhance the performance of Transformer-based classifiers on appliance detection tasks. 


In this paper, we propose an Appliance Detection Framework (ADF) designed to detect the presence/absence of appliances in long and variable length consumption time series.
The ADF framework takes as input the \emph{subsequences} of a client consumption series.
In the end, the prediction probabilities made on each subsequence are merged, allowing the model to predict the final detection label for the entire series.
We also introduce TransApp, a deep-learning Transformer-based classifier that is first trained in a self-supervised manner using non-labeled consumption time series.
Then, the model is fine-tuned on labeled data to detect a specific appliance in a consumption series. 
Overall, TransApp is more accurate than all state-of-the-art competitors, and the ADF framework renders it scalable to the sizes of real-world datasets.
In summary:

    
    \noindent$\bullet$ We propose the Appliance Detection Framework (ADF) to detect the presence of appliances in households, using real-world consumption series, which are sampled at a very low frequency, and are long and variable-length.
    ADF addresses these challenges by operating at individual subsequences of each consumption series, instead of each series in its entirety.
    
    \noindent$\bullet$ We propose TransApp, a Transformer-based time series classifier, which can first be pretrained in a self-supervised manner to enhance its ability on appliances detection tasks.
    This way, TransApp can significantly improve its accuracy.
    
    \noindent$\bullet$ We 
    evaluate our proposed approach (ADF + TransApp) in terms of accuracy and scalability on two real-world datasets, and make the code publicly available~\cite{transappcode}.
    %
    The results demonstrate the superiority of our solution against previous approaches for appliance detection, including state-of-the-art time series classifiers. 
    
    \noindent$\bullet$ Finally, we highlight the benefit of the self-supervised training process on non-labeled consumption series so as to enhance the performance of TransApp on appliance detection tasks.

\section{Related work \& Problem Definition}

\subsection{Appliance Detection}

Appliance detection is a problem related to Non-Intrusive Load Monitoring (NILM), which aims at identifying the power consumption, pattern, or on/off state activation of individual appliances using only the total consumption series~\cite{reviewnilm}.
Even though detecting an appliance can be seen as a step of NILM-based methods~\cite{QU2023112749, ASLAN2022112087, 7498597, 6583496, 10.1145/3077839.3077845, https://doi.org/10.48550/arxiv.2209.03759, huquet_activation}, 
they differ from our objective for two main reasons. 
First, the vast majority of NILM studies relied on smart meter data recorded at 1Hz (or more), which is much more detailed than the datasets available in practice. 
Moreover, they need either knowledge about how each appliance operates, or training on their individual power consumption patterns.
Second. these studies essentially focus on detecting \emph{when} a specific appliance is "ON" rather than \emph{if} a household owns a specific appliance. 
In addition, the presence of a specific appliance is already known before applying these approaches.

Nevertheless, few studies exist in the literature that try to detect if a specific appliance is present in a household, using consumption time series sampled at very low-frequency (one point collected every 1min or more)~\cite{Albert_2013, dengetal, appliancedetectionbenchmark}. 
In~\cite{Albert_2013}, the authors use a Hidden Semi-Markov Model (HSMM) to extract features from the consumption series of a house. 
These features are then merged with external variables (such as temperature) and serve to train an AdaBoost classifier~\cite{adaboost} to detect users' characteristics, such as the presence of appliances.
The study shows encouraging results in identifying some appliances, such as Electric Dryers or Washing Machines. 

Deng et al.~\cite{dengetal} proposed a framework to detect the presence of appliances in a consumption series based on subsequences, and using a modified ResNet classifier (cf. Section~\ref{subsec:TSC}). 
This approach first addresses the issue of unbalanced classes by employing an oversampling slicing approach to obtain equal numbers of training class instances. 
Then, it takes subsequences of the consumption series as input, and uses majority voting with a threshold parameter to predict the final label. 
However, the evaluation of the approach 
does not measure the overall performance in detecting the presence/absence of appliances, and the study compares the proposed approach to an HSMM-based classifier, but not against the state-of-the-art time series classifiers for appliance detection.

A recent study~\cite{appliancedetectionbenchmark} proposes an extensive benchmark of state-of-the-art time series classifiers applied to the appliance detection problem, using several very low-frequency electricity consumption datasets.
The results show that the convolutional-based time series classifiers (random kernel and deep learning based) outperform other approaches regarding detection performance, and are the most scalable to the length of the series.
However, the accuracy results obtained using 30min sampled data 
are far from perfect for several appliance detection cases, showing that more work is needed in this area.

\subsection{Time Series Classifiers}
\label{subsec:TSC}

Time series classification (TSC) \cite{https://doi.org/10.48550/arxiv.1602.01711, 10.1007/s10618-019-00619-1} is an important analysis task across several domains.
Many studies have suggested different approaches to solving the TSC problem, such as comparing similarity measures between time series~\cite{1053964}, identifying discriminant patterns~\cite{Hills2014}, detecting rare patterns (e.g., anomalies)~\cite{10.1145/3514221.3526183}, and addressing class imbalance~\cite{ideals}. 
In addition, benchmarks, such as the UCR archive~\cite{UCR2018}, have been proposed, on which exhaustive experimental studies have been conducted~\cite{https://doi.org/10.48550/arxiv.1602.01711}.
In the following paragraphs, we provide a brief overview of the state-of-the-art classifiers, selected based on their performance in previous studies and their suitability for the appliance detection tasks~\cite{10.1145/3514221.3526183, appliancedetectionbenchmark}.

The RandOm Convolutional KErnel Transform (ROCKET) algorithm utilizes~\cite{DBLP:journals/corr/abs-1910-13051} random convolutional generated kernels to extract features from raw time series. 
It generates a set of random filters and extracts maximum values and positive value proportions as new features for classification using a ridge classifier.
Arsenal~\cite{hivecote2} is an ensemble of multiple ROCKET classifiers that uses a restricted number of kernels while estimating classification probabilities without modifying the ridge classifier.

Convolutional Neural Network (CNN)~\cite{DBLP:journals/corr/OSheaN15} is a deep learning architecture commonly used in image recognition. 
The ConvNet variant~\cite{https://doi.org/10.48550/arxiv.1611.06455}, we use in this study employs stacked convolutional blocks with specific kernel sizes and filters, followed by global average pooling and linear layers for classification.
The Residual Network (ResNet) architecture~\cite{https://doi.org/10.48550/arxiv.1512.03385} addresses the gradient vanishing problem in large CNNs~\cite{simonyan2015a}. 
The adaptation for time series classification~\cite{https://doi.org/10.48550/arxiv.1611.06455} consists of stacked residual blocks with residual connections, where each block contains 1D convolutional layers with the same kernel sizes and filters. 
A global average pooling~\cite{https://doi.org/10.48550/arxiv.1312.4400}, a linear layer, and a softmax activation are used for classification.
The ResNet architecture was extended to utilize dilated convolutions to increase the receptive field~\cite{dengetal}, also adding two encoder/decoder modules after the convolutional block with a dot product attention mechanism (hence, called ResNet with Attention Mechanism). 
After feature extraction, classification is performed using a multi-layer perceptron and a softmax activation.
Inspired by inception-based networks~\cite{inception2014} for image classification, InceptionTime~\cite{Ismail_Fawaz_2020} is designed for time series classification.
It employs Inception modules composed of concatenated convolutional layers using different filter sizes.
The outputs are passed through activation and normalization layers, at the end, classification is performed using a global average pooling, followed by a linear layer and softmax activation function.

The Transformer architecture 
has demonstrated remarkable success in various tasks~\cite{bert, gpt}, 
and has also been investigated for time series analysis, demonstrating promising results on downstream tasks such as forecasting, anomaly detection, and classification~\cite{choosewisely, tst_survey, informer, autoformer, fedformer, Yuqietal-2023-PatchTST, zerveas_tst}.
Nevertheless, the attention mechanisms used in the Transformer~\cite{attentionisallyouneed} has a complexity of $\mathcal{O}(N^2)$ on both time and space (with $N$ the length of the input sequence), making this type of model not scalable for long time series (more than 10000 points).
In the energy domain, two recent studies investigated the use of Transformer-based architecture to perform energy disaggregation using second sampled data and achieve state-of-the-art performance against other deep learning architectures~\cite{bert4nilm, electricitynilm}.

\subsection{Problem Formulation}
\label{subsec:pbformulation}


An electrical consumption time series is defined as a univariate time series $X=(x_1, ..., x_T)$ of ordered elements $x_j \in \mathcal{R}_{+}^{1}$ following $(i_1,...,i_T)$ time consumption indexes (i.e., timestamps). 
Furthermore, we refer in the study to \emph{very low frequency} consumption time series for data sampled at more than 1min.

\noindent{\textbf{[Appliance Detection Problem]}} In this work, we treat the appliance detection problem as a supervised binary classification problem. 
According to a collection of consumption time series $\mathcal{X} = (X^1, ..., X^N$), which can be of variable-length, we want to predict the presence/absence of an appliance $a$ in a time series $X^i$.

\section{The ADF}
\label{sec:approach}

\begin{figure*}
    \centering
     \includegraphics[width=0.8\linewidth]{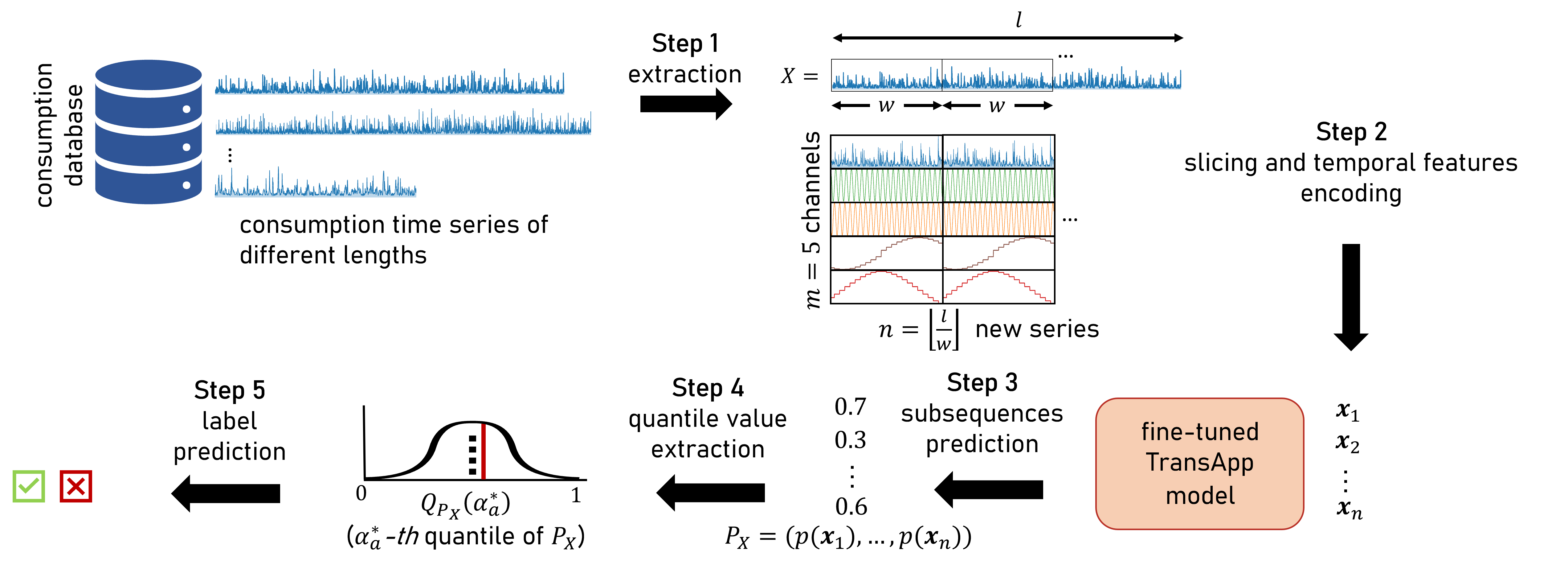}
     \vspace*{-0.4cm}
    \caption{\label{fig:Framework} Overview of our proposed Appliance Detection Framework (ADF).}
     \vspace*{-0.2cm}
\end{figure*}


\label{subsec:Framework}

We propose ADF, an Appliance Detection Framework that takes fragments of an entire consumption time series as input.
Subsequently, the classifier used inside the framework is trained using subsequences.
The proposed framework (see Figure~\ref{fig:Framework}) uses the following steps to detect the presence of an appliance $a$ in a consumption time series $X$.

    \noindent\textbf{Step 1.} First, we extract from a database a client's consumption time series $X$ of length $l$. 

    \noindent\textbf{Step 2.} The time series $X$ is then sliced in $n$ new non-overlapping subsequences of length $w$, using a tumbling window.
    In addition, to keep positional information about the time of the days and hours, we concatenate the sliced sequences with temporal encoded features (see Section~\ref{subsubsec:cyclicalfeatures}.a for details).
    This results in $n = \lfloor \frac{l}{w} \rfloor$ new multivariate time series $\boldsymbol{x}^{w \times m}$. 
    With $m$, the number of channels.
    
    \noindent\textbf{Step 3.} Afterward, we feed each subsequence to a TransApp classifier instance previously finetuned to detect the specific appliance $a$.
    The model then predicts a given detection probability for each subsequence $\boldsymbol{x}_i$, resulting in a vector of probability $P_X = (p(\boldsymbol{x}_1), ..., p(\boldsymbol{x}_n))$.
    
    \noindent\textbf{Step 4.} The value of the $\alpha^{*}_{a}$-th quantile is then extracted from $P_X$, allowing us to obtain the final prediction probability of our model according to the entire consumption series (we discuss the tuning of the $\alpha^{*}_{a}$ parameter in the following Section~\ref{subsubsec:quantile}.b).
    
    \noindent\textbf{Step 5.} At the end, the final predicted label is given by rounding the extracted value.

In addition, we note that this general framework can be used with any classifier able to predict detection probabilities.

\paragraph{a) Temporal Features Encoding}
\label{subsubsec:cyclicalfeatures}

The usage of various appliances often correlates with specific times of the day (e.g., electric cookers used during mealtime).
However, the slicing process used in our framework leads to subsequences that start and end at different time points.
To improve the model's understanding of the correlation between these time-related patterns and enhance detection, we introduce additional channels 
to encode time.
More precisely, we add new channels as encoded features related to the days and hours by projecting these discrete features on a $Sin/Cos$ basis:
\begin{equation*}
    Te_{sin}(i_t) = \sin \left(\frac{2 \pi i_t}{p} \right) \text{ and }
    Te_{cos}(i_t) = \cos\left(\frac{2 \pi i_t}{p} \right),
\end{equation*}
with $i_t=\{1,...,24\}$ and $p = 24$ for hour encoding, and $i_t=\{1,...,7\}$ and $p = 7$ for days encoding.
We motivate the choice of adding these new channels instead using only the univariate consumption series in the ablation study conducted in Section~\ref{sec:AblationStudy}.

\paragraph{b) Quantile Parameters Tuning}
\label{subsubsec:quantile}

As the model predicts a probability for each subsequence $\boldsymbol{x}_i$, we need to merge these values to assign the final label to the entire consumption series $X$.
The most straightforward approach could be to take the mean of these probabilities (majority voting).
However, this includes a too-strong hypothesis that the appliance is present uniformly in each sub-sequence $(\boldsymbol{x}^{w \times m}_1, ..., \boldsymbol{x}^{w \times m}_n)$.
Thus, for an appliance $a$ and a predicted vector of discrete probabilities $P_X = (p(\boldsymbol{x}_1), ..., p(\boldsymbol{x}_n))$, we extract the values corresponding to its $\alpha^{*}_{a}$-th quantile (defined as $Q_{P_X}(\alpha^{*}_{a})$).
The $\alpha^{*}_{a}$ parameter is defined according to a validation dataset during the training process of our framework.
More precisely, for an appliance $a$ we compute the classification measure for each extracted values $Q_{P_X}(\alpha_{a})$ (with $\alpha \in \{0,0.5, ..., 0.95,  1\}$) and keep the one that maximizes the chosen score (F1-Macro Score in our experiments).
Formally, for an appliance $a$, a validation dataset $\mathcal{D}_v = \{X^1, ..., X^N \}$, and an accuracy measure $S$, we define the parameters as follows:
    $\alpha^{*}_{a} = \underset{\alpha \in \{0,0.5, ..., 0.95,  1\}}\argmax~S \left(y_{true}, y^{\alpha}_{pred}\right)$,
with $y^{\alpha}_{pred} = (Q_{P_{X^1}}(\alpha), ..., Q_{P_{X^N}}(\alpha))$ and $y_{true}$ ground true label for each consumption series.
Note that this $\alpha^{*}_{a}$ parameter is reminiscent of the tuned threshold voter parameter in Deng et al.~\cite{dengetal}. 
However, using a parameter tuned according to the \emph{predicted probabilities} (instead of the \emph{predicted label}) on each subsequence allows us to consider the model's confidence on each one of them.

\subsection{TransApp Architecture}
\label{subsec:transapparchitecture}

For the past few years, Convolutional Neural Networks have shown great ability to extract discriminative patterns for Time Series Classification~\cite{Ismail_Fawaz_2020, DBLP:journals/corr/abs-1910-13051}, and more specifically, great performance when applied to detect appliance in consumption time series~\cite{appliancedetectionbenchmark}.
On the other hand, Transformer architectures are known to be good candidates for learning universal representation of data and perform well on several applications~\cite {bert, vit, vitmae, tst_survey}. 
However, as mentioned in recent studies, the vanilla transformer architecture is not designed initially to treat time series data~\cite{Yuqietal-2023-PatchTST}.
Indeed, a single-time step value of a data series does not have value by itself.
Therefore, inspired by similar architecture proposed for image classification~\cite{coatnet}, we propose TransApp as a deep learning time series classifier that combines these two types of architecture: TransApp combines a robust embedding block based on dilated convolutions with a Transformer block.
The convolutional block serves as a features extractor to give an inductive bias of localized patterns and helps the model perform better on classification tasks.
The Transformer module learns long-range dependencies and is a key part of our architecture to extract representation and benefit of our pretraining process.

\begin{figure*}
    \centering
     \includegraphics[width=1\linewidth]{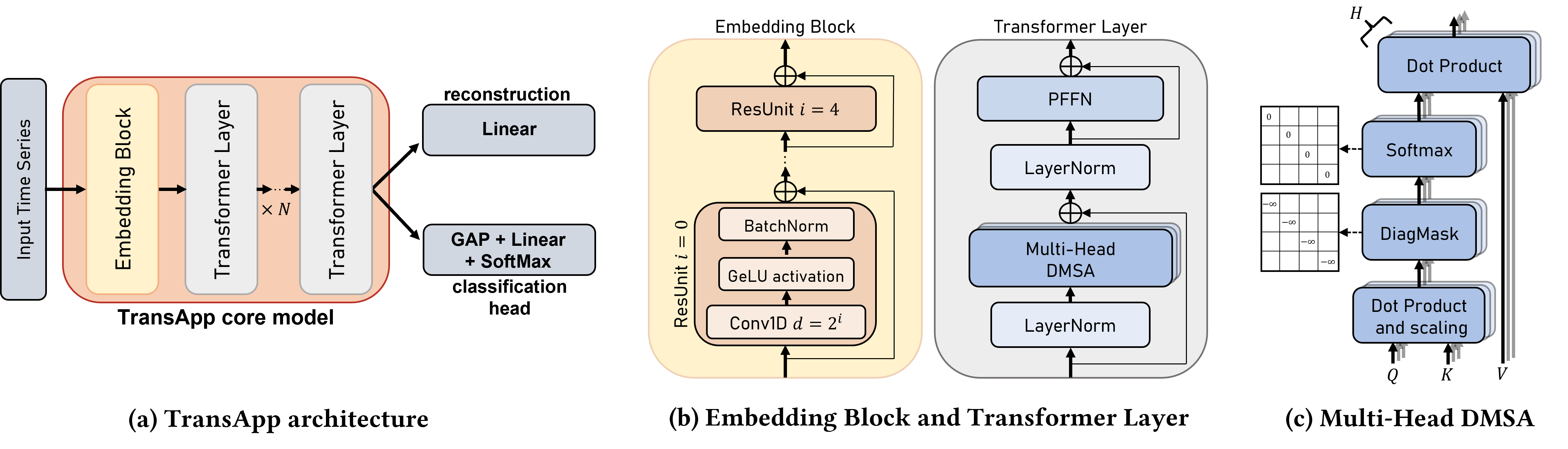}
     \vspace*{-0.8cm}
    \caption{\label{fig:transappmodel} Overview of the TransApp architecture.}
\end{figure*}

The TransApp core model, shown in Figure~\ref{fig:transappmodel}(a), results in an encoder that maps an input series $\boldsymbol{x}^{w \times m}$ to a latent space $\boldsymbol{z}^{w \times d_{\text{model}}}$, where $w$ is the length and $m$ the number of variables of the input time series, and $d_{\text{model}}$ the dimension of the latent space, a.k.a. inner model dimension ($d_{\text{model}}=96$ in our experiments).
The latent representation $z^{w \times d_{model}}$ is used by a specific head to perform the pretraining (sort of denoising), or classification task. 



\subsubsection{Embedding Block}

The first block of our model, shown in Figure~\ref{fig:transappmodel}(b), results in 4 stacked convolutional Residual Units (ResUnit).
Each ResUnit is composed of a convolutional layer, a GeLU activation function~\cite{gelu}, and a BatchNormalization layer~\cite{https://doi.org/10.48550/arxiv.1502.03167}.
For each ResUnit $i=1,...,4$, a dilation parameter $d=2^i$ that exponentially increases according to the ResUnit's depth is employed. 
It has been experimentally proven to significantly broaden the receptive fields for time series applications, in contrast to using a constant dilation parameter~\cite{https://doi.org/10.48550/arxiv.1803.01271}. 
A stride parameter of 1 is employed to keep the time dimension identical to the input series.
Furthermore, a residual connection is used between each Residual Unit to ensure stability during the training.

\subsubsection{Transformer Block}
\label{subsec:transencoderblock}

The second block of our model, visible in Figure~\ref{fig:transappmodel}(b), results in $N$ stacked Transformer layers ($N$ set to 3 or 5 in our experiments) using pre-Layer Normalization.
Each Transformer layer comprises the following elements: a layer normalization, a Multi-Head Diagonally Masked Self-Attention mechanism (Multi-Head DMSA), a second layer normalization, and a Positional Feed-Forward Network~\cite{attentionisallyouneed} (PFFN). 
We introduce residual connections after the Multi-Head DMSA and the PFFN and the use of a Dropout parameter to prevent overfitting.

Note that we do not apply any positional encoding before the Transformer block as it is usually used in most of the Transformer architectures~\cite{attentionisallyouneed, vit, tst_survey, zerveas_tst}.
Our experiments in Section~\ref{sec:AblationStudy} demonstrate that using fixed or fully learnable positional encoding leads to a deterioration in results.
We assume this is because the multivariate sequences given at input already incorporate temporal encoded features, rendering positional encoding unnecessary.

\noindent{\textbf{[Diagonally Masked Self-Attention]}} In our Transformer block, we utilize a modified version of the original Self-Attention mechanism~\cite{attentionisallyouneed} called Diagonally Masked Self-Attention (DMSA). 
DMSA involves applying a mask to the diagonal elements of the attention score matrix, forcing the scores to be zero after the softmax operation. 
This modification emphasizes inter-token relations and enhances the model's ability to capture meaningful dependencies.
In our architecture, we used Multi-Head DMSA, a multi-head implementation of DMSA similar to the original attention mechanism~\cite{attentionisallyouneed} that uses $H$ different parallel projection sets.

We motivate the choice of DMSA against the original mechanism in the ablation study conducted in Section~\ref{sec:AblationStudy}.
Note that this type of diagonal attention mask has already proven better performance against the original one in Vision Transformer~\cite{improvingvitsmalldataset} to improve the overall performance of the model when dealing with small-size dataset, and more recently for time series imputation~\cite{saits}.

\subsection{Two-step training process}

The training of our proposed TransApp architecture results in a two-step training, illustrated in Figure~\ref{fig:twosteptraining_overview}.

\subsubsection{Self-supervised Pretraining}
\label{subsec:selfsupervised}

\begin{figure}
    \centering
     \includegraphics[width=1\linewidth]{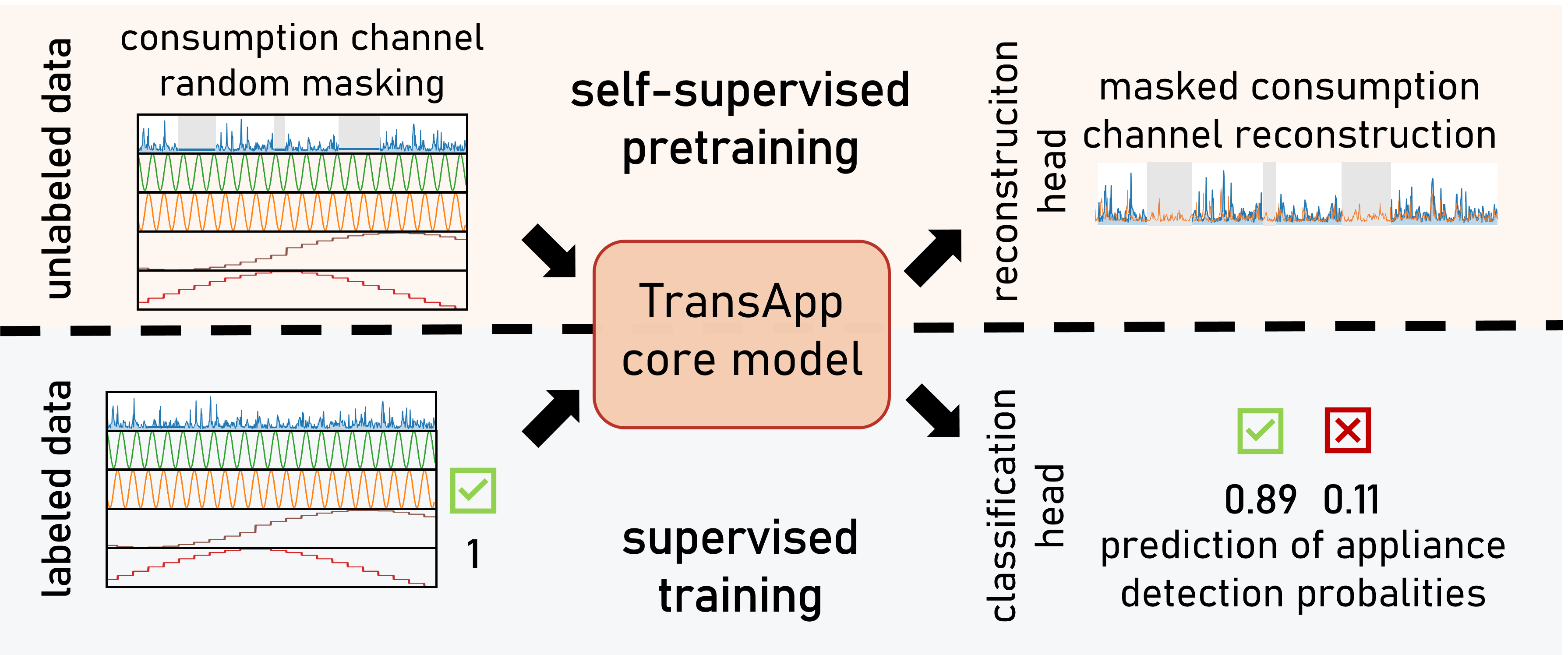}
     \vspace*{-0.6cm}
    \caption{Overview of the TransApp two steps training.}
    \label{fig:twosteptraining_overview}
     \vspace*{-0.4cm}
\end{figure}

The use of a self-supervised pretraining of a Transformer architecture on an auxiliary task has been used in the past 
to boost the model performance on downstream tasks~\cite{bert, vitmae, zerveas_tst, Yuqietal-2023-PatchTST, electricitynilm}.
This process is inspired by the mask-based pretraining of vision transformer~\cite{vitmae}.
As shown in Figure~\ref{fig:twosteptraining_overview}, it requires only the input consumption series without any appliance information label.
The training results in a reconstruction objective of a corrupted (masked) time series fed to the model input.
To do that, as shown in Figure~\ref{fig:transappmodel}(a), we use a reconstruction head after the core model architecture, resulting in a linear layer that maps for each time step the latent representation obtained after the Transformer layers $z^{win \times d_{\text{model}}}$ to a space $\hat{x}^{win \times 1}$.

\noindent{\textbf{[Masking Process]}} The masking process used in our approach is inspired by the one proposed in~\cite{zerveas_tst}. 
It aims to introduce random corrupted segments inside the input sequence (i.e., values set to 0).
However, as shown in Figure~\ref{fig:twosteptraining_overview}, we apply the mask only on 
the consumption series channel, leaving the temporal encoded feature channels untouched.
Furthermore, in our experiments, we use a high masking ratio corresponding to 50\% of the original time series masked with segments of length $l_m=24$ on average (this length corresponds to periods of 12h masked values in average using 30min sampled data).

\noindent{\textbf{[Reconstruction Loss]}} During the self-supervised process, the model is trained using a Loss function that calculates the Mean Absolute Error between the predicted and true values of the masked elements of an input consumption time series. It is formally defined as :
    $\mathcal{L}_{MAE} = \frac{1}{\#M}\sum_{i}^{}|\hat{x}_i - x_i|_{1}$, 
with $\#M$ the number of masked elements in the input series.

\subsubsection{Supervised Training}
\label{subsec:finetuning}

As depicted in Figure~\ref{fig:twosteptraining_overview}, the supervised training results in a simple binary classification process using labeled time series.
In addition, we used a classification head for this step (see Figure~\ref{fig:transappmodel}(a)).
This head comprises a global average pooling \cite{globalaveragepooling} applied along the sequence length dimension, followed by a linear layer that maps the results to the number of classes (2 in our case).
Note that the label of the entire consumption series is assigned to all sliced subsequences during the training process.

\section{Experimental Evaluation}
\label{sec:exp}
 
All experiments are performed on a high-performance computing cluster node with 2 Intel Xeon Gold 6140 CPUs with 190 Go RAM and 2 NVidia V100 GPUs with 16Go RAM. 
The source code~\cite{transappcode} of our framework is in Python 3.7, and the core of our model is implemented using the 1.8.1 version of PyTorch~\cite{https://doi.org/10.48550/arxiv.1912.01703}.

\subsection{Datasets}
\label{subsec:datasets}

\begin{table}[tb]
\caption{\label{table:detaildataset} Dataset characteristics: number of time series $\sharp$TS, number of label instances $\sharp$Labels 0/1, and class imbalance.}
{\scriptsize
\begin{tabular}{c|c|c|c|c}
    \toprule
    \multicolumn{2}{c|}{\textbf{Detection cases}} & \multicolumn{3}{c}{\textbf{Dataset description}} \\
    Datasets & Appliances & $\sharp$TS &  $\sharp$Labels 0/1 &  Imbalance Ratio  \\
    \midrule
    \multirow{9}{*}{CER} & Desktop Computer & 3470 & 1827/1643 & 0.47  \\
    & TVs (greater 21") & 3470 & 540/2930 & 0.84 \\
    & TVs (less 21") & 3470 & 1212/2258 & 0.65 \\
    & Laptop Computer & 3470 & 1618/1852 & 0.53 \\
    & Cooker & 3482 & 841/2641 & 0.76  \\
    & Dishwasher & 3482 & 1175/2307 & 0.66 \\
    & Tumble Dryer & 3482 & 1107/2375 & 0.68 \\
    & Water Heater & 3488 & 1535/1953 & 0.56 \\
    & Plugin Heater & 3482 & 2394/1088 & 0.31 \\
    \hline
    \multirow{7}{*}{EDF 1} & Tumble Dryer & 3365 & 2071/1294 & 0.38 \\
    & Dishwasher & 3372 & 279/3093 & 0.93 \\
    & Convector/HeatPump & 1478 & 468/1010 & 0.68 \\
    & Air Conditioner & 3388 & 524/2864 & 0.85 \\
    & Water Heater & 4685 & 1635/3050 & 0.53 \\
    & Heater & 4685 & 2202/2483 & 0.53 \\
    & Electric Vehicle & 558 & 330/228 & 0.41 \\
    
    \bottomrule
\end{tabular}
}
\end{table}

We now describe the datasets we used in this study (see Table~\ref{table:detaildataset}).
The first one comes from the Irish Social Science Data Archive (ISSDA)~\cite{ISSDA}, and the two other two are from Electricité De France (EDF), the main French electricity supplier.  

\noindent{\textbf{[CER Dataset]}} The Commission for Energy Regulation of Ireland conducted a study between 2009-2011 to assess the performance of smart meters and their impact on consumer energy consumption~\cite{CER_2012}. 
Over 5000 Irish homes and businesses were monitored with communicant Smart Meters, recording the electricity consumption every 30min. 
For our experiments, we were only interested in the residential sub-group of the study, i.e., 4225 households recorded from July 15, 2009, to January 1, 2011. 
This sample results in 4225 electricity consumption series of length 25728 each.
For the study, the participants filled out a questionnaire allowing us to know the household composition and appliances present in the household.

\noindent{\textbf{[EDF Dataset]}} As part of these activities, EDF collects data from customers who give their consent.

\noindent{\textbf{$\bullet$ EDF 1}}
This dataset is based on a survey conducted by EDF to better understand its customers and their electricity consumption behavior.  
The total power consumption of 4701 houses was recorded every 30min. 
The dataset contains variable-length consumption series collected between September 2018 and December 2022.
The dataset's time series have an average length of 21630 points (min: 1338, max: 40300).
Like the CER dataset, customers filled out a questionnaire with information about household composition, including the appliances in the house.

\noindent{\textbf{$\bullet$ EDF 2}}
The second dataset provided by EDF results in a huge collection of nearly 200 000 variable-length consumption series recorded from different households.
However, this dataset only provides the house's power consumption recorded every 30min, without any information about the appliances present in the house.

\begin{table*}
\caption{Results (average Macro F1-score for 3 runs) for the different approaches (time series classifier, frameworks) applied to the detection cases using the CER dataset. The best score is shown in bold, and the second best is underlined.}
\label{table:CERresults}
\begin{adjustbox}{width=\textwidth,center}
\begin{tabular}{c|ccccc|c|cccccccccc}
\toprule
     & \multicolumn{15}{c}{\textbf{Models comparaison}} \\
     & \multicolumn{5}{c|}{\textbf{Time Series Classifiers}} & \textbf{Deng et al.} & \multicolumn{9}{c}{\textbf{ADF with}} \\
     \textbf{Appliance} & Arsenal & ROCKET & ConvNet & ResNet & Inception & \textbf{framework} & Arsenal & ConvNet & ResNet & Inception & Informer & Autoformer & Fedformer & TransApp & TransAppPT \\
     \textbf{Detection Cases} &  &  &  &  &  & $w=512$ & $w=1024$ & $w=512$ & $w=512$ & $w=1024$ & $w=256$ & $w=256$ & $w=512$ & $w=1024$ & $w=1024$ \\
\midrule
    Cooker & 0.68 & 0.676 &  0.665 & 0.698 & 0.705 & 0.670 & 0.723 & 0.71 & 0.732 & 0.729 & 0.71 & 0.704 & 0.62 & \underline{0.746} & \textbf{0.754} \\
    
    Dishwasher & 0.705 & 0.708 & 0.736 & 0.72 & 0.732 & 0.652 & 0.726 & 0.719 & 0.716 & \textbf{0.758} & 0.644 & 0.661 & 0.653 & \underline{0.742} & 0.738 \\
    
    Water Heater & \textbf{0.631} & 0.616 & 0.615 & 0.617 & \textbf{0.631} & 0.519 & \underline{0.629} & 0.608 & 0.609 & 0.614 & 0.581 & 0.598 & 0.479 & 0.601 & 0.616 \\
    
    Plugin Heater & 0.509 & 0.507 & 0.539 & 0.497 & 0.556 & 0.536 & \underline{0.609} & 0.597 & 0.581 & 0.568 & 0.540 & 0.563 & 0.519 & 0.589 & \textbf{0.624} \\
    
    Tumble Dryer & 0.645 & 0.641 & 0.605 & 0.589 & 0.616 & 0.615 & \textbf{0.657} & 0.644 & 0.658 & 0.656 & 0.619 & 0.624 & 0.507 & 0.654 & \underline{0.655} \\
    
    TVs (greater 21") & 0.555 & 0.55 & 0.58 & 0.589 & 0.591 & 0.576 &  \underline{0.592} & 0.579 & 0.577 & \underline{0.592} & 0.579 & 0.581 & 0.534 & 0.578 & \textbf{0.593} \\
    
    TVs (less 21") & 0.539 & 0.526 & 0.449 & 0.454 & 0.513 & 0.519 & 0.54 & 0.537 & \textbf{0.549} & 0.544 & \underline{0.545} & 0.534 & 0.463 & 0.52 & 0.543 \\

    Desktop Computer & 0.618 & 0.609 & 0.606 & 0.61 & 0.609 & 0.537 & \underline{0.624} & 0.601 & 0.608 & 0.59 & 0.605 & 0.595 & 0.566 & \textbf{0.626} & 0.612 \\
    
    Laptop Computer & 0.641 & 0.632 & 0.621 & 0.64 & 0.620 & 0.573 & 0.642 & 0.637 & 0.64 & 0.624 & 0.617 & 0.636 & 0.538 & 0.64 & \textbf{0.652} \\
    \hline
    \textit{Avg. Score} & 0.614 & 0.604 & 0.599 & 0.597 & 0.619 & 0.577 & \underline{0.638} & 0.626 & 0.63 & 0.631 & 0.604 & 0.61 & 0.542 & 0.633 & \textbf{0.643}\\
    \textit{Avg. Rank} &  7.5 &  9.778 & 10 & 8.778 & 7.556 & 12.889 & \underline{3.278} & 7.556 & 5.778 & 5.611 & 9.667 & 9.222 & 14.111 & 5.556 & \textbf{2.722} \\
\bottomrule
\end{tabular}
\end{adjustbox}
\end{table*}

\begin{table*}
\caption{Results (average Macro F1-score for 3 runs) for the different approaches applied to appliance detection cases using the EDF 1 dataset. 
The best score is shown in bold, and the second best is underlined.}
\label{table:EDFResults}
\begin{adjustbox}{width=\textwidth,center}
\begin{tabular}{c|ccccc|c|cccccccccc}
\toprule
     & \multicolumn{16}{c}{\textbf{Models comparaison}} \\
     & \multicolumn{5}{c|}{\textbf{Time Series Classifiers}} & \textbf{Deng et al.} & \multicolumn{10}{c}{\textbf{ADF with}} \\
    \textbf{Appliance} & Arsenal & ROCKET & ConvNet & ResNet & Inception  &  \textbf{framework} & Arsenal & ConvNet & ResNet & Inception & Informer & Autoformer & Fedformer & TransApp & TransAppPT (EDF 1) & TransAppPT-l (EDF 2) \\
    \textbf{Detection Cases} &  &  &  &  &  & $w=256$ & $w=1024$ & $w=1024$ & $w=1024$ & $w=1024$ & $w=256$ & $w=512$ & $w=256$ & $w=1024$ & $w=1024$ & $w=1024$ \\
    \midrule
    Electric Heater & 0.808 & 0.817
& 0.736 & 0.748 & 0.757 & 0.793 & 0.815 & 0.796 & 0.800 & 0.805 & 0.77 & 0.725 & 0.731 & 0.815 &  \underline{0.818} & \textbf{0.828} \\
    
    Convector/HeatPump & 0.678 & 0.674 & 0.605 & 0.542 & 0.578 & 0.659 & \underline{0.711} & 0.689 & 0.706 & 0.701 & 0.652 & 0.563 & 0.552 & 0.706 &  0.706 &  \textbf{0.736} \\

    Air Conditioner & 0.543 & 0.542 & 0.531 & 0.494 & 0.528 & 0.610 & 0.577 & 0.603 & 0.568 & 0.623 & 0.598 & 0.558 & 0.548 & 0.656 & \underline{0.669} & \textbf{0.67} \\
    
    Water Heater & 0.827 & 0.825 & 0.774 & 0.814 & 0.828 & 0.836 & 0.835 & 0.822 & 0.834 & 0.840 & 0.79 & 0.773 & 0.753 & 0.844 & \underline{0.849} & \textbf{0.855} \\

    Dishwasher & 0.514 & 0.523 & 0.578 & 0.581 & 0.568 & 0.560 & 0.577 & 0.553 & 0.571 & 0.555 & 0.521 & 0.505 & 0.5 & 0.564 & \underline{0.594} & \textbf{0.601} \\
    
    Tumble Dryer & 0.674 & 0.675 & 0.595 & 0.642 & 0.625 & 0.635 & 0.698 & 0.662 & 0.685 & 0.701 & 0.570 & 0.584 & 0.591 & 0.694 & \underline{0.708} & \textbf{0.709} \\

    Electric Vehicle & 0.7 & 0.709 & 0.626 & 0.682 & 0.693 & 0.726 & 0.757 & 0.707 & 0.750 & 0.778 & 0.709 & 0.683 & 0.661 & 0.782 & \underline{0.807} & \textbf{0.825} \\

    \midrule
    \textit{Avg. Score} & 0.678 & 0.681 & 0.635 & 0.643 & 0.654 & 0.688 & 0.719 & 0.69 & 0.70 & 0.715 & 0.633 & 0.627 & 0.619 & 0.723 &  \underline{0.736} & \textbf{0.746} \\
    \textit{Avg. Rank} & 9.714 & 8.929 & 12.429 & 12 & 11.286 & 8.143 & 4.929 & 9 & 6.571 & 5.429 & 11.357 & 14 & 14.571 & 4.357 & \underline{2.286} & \textbf{1} \\
\bottomrule
\end{tabular}
\end{adjustbox}
\end{table*}

\subsection{Evaluation Procedure}
\label{subsec:evalprocedure}

We use different types of baselines to compare to our TransApp architecture.
First, we include the framework proposed by Deng et al.~\cite{dengetal}.
Then, according to the results of~\cite{appliancedetectionbenchmark}, we selected the 5 best time series classifiers for the appliance detection problem (Arsenal, ROCKET, ConvNet, ResNet, and Inception).
Finally, 
we include 3 additional state-of-the-art Transformer-based models proposed for time series modeling (Informer, Autoformer, and Fedformer).

For the two datasets, we evaluated the effectiveness of ADF by comparing the performance of selected state-of-the-art classifiers inside and outside ADF.
For all the baselines, we use the same 70\%/10\%/20\% random split of the dataset for the training, validation, and test sets.
The validation set stops the training and prevents overfitting using deep-learning classifiers.
Furthermore, it is used for our framework and the one proposed by Deng et al.~\cite{dengetal} to calculate the optimal $\alpha^{*}_{a}$ and voter threshold parameter of the respective merging process.
We balance the data before training for all the tested classifiers: we apply the subsequences oversampling method proposed in the original paper when using the framework proposed by Deng et al.~\cite{dengetal}, and we equalize each class using a random undersampling process for all the other approaches.

\noindent{\textbf{[Outside ADF]}} Since the consumption series of the EDF 1 dataset are of different lengths, we used standard padding methods to train the classifier outside ADF with this dataset (by setting all the time series to a specified maximum length).
More specifically, we pad the time series with 0 at the end to meet the length of the longest time series of the dataset.
For the deep learning-based methods, we also tried to train the classifier using a batch of 1 (as convolutional-based methods are insensitive to time series length); however, this led to poorer results than padding the series to equal length.
Note that Informer, Autoformer, Fedformer, and our TransApp architecture could not train outside ADF (using as input the entire series of length 25728 for the CER dataset and 40300 for the EDF 1 dataset) due to GPU memory issues (even with a batch size of 1).

\noindent{\textbf{[Inside ADF]}} We evaluate ADF coupled with all the baselines with different subsequences windows length $w$ to assess the sensitivity of this parameter on the results of each classifier.
We performed experiments with $w = \{256, 512, 1024, 2048, 4096\}$ for the CER dataset. 
For the EDF 1 dataset, we used only $w = \{256, 512, 1024\}$ as the shortest time series has length 1338 in this dataset.
We also note that Rocket was not evaluated inside ADF as this classifier is not designed to predict probabilities.
For the pretraining part of our proposed TransAppPT model, we use all available time series without considering any label information. 
We named TransApp the no-pretrained architecture and TransAppPT the pretrained one.
In addition, we also evaluate a TransApp architecture pretrained on the EDF 2 dataset that is then finetuned to the EDF 1 dataset (TransAppPT-l, composed of 5 Transformer layers instead of 3)\footnote{Pretraining TransApp with 3 Transformer layers on the EDF 2 dataset leads to equivalent results as pretraining it on EDF 1. Thus, we investigate a larger architecture that can benefit from this large dataset.}.

\subsubsection{Accuracy Measure and Addressing Imbalanced Datasets}
\label{subsec:measures}

Accuracy 
represents the fraction of correct predictions across all classes. 
However, it treats each class equally, regardless of class distribution, making it inadequate for imbalanced datasets.
Precision, recall, and the F1-score are standard measures to address this issue. 
Precision (Pr) is the percentage of class instances correctly classified, while recall (Rc) is the percentage of misclassified instances. 
The F1-score is the harmonic average of precision and recall, providing a balanced measure.
For binary classification problems with imbalanced data, these measures are often applied only to the minority class. 
However, in appliance detection scenarios, the minority class may vary. 
Therefore, to evaluate overall performance and account for variability, the Macro F1-score is used, which calculates the average F1-score across all classes: 
$\text{Macro F1-score} = \frac{1}{N} \sum_{i=1}^{N} \text{F1-score}_i$, 
where $N$ is the number of classes (equal to 2 in this study).

\subsubsection{Scalability} We evaluate the scalability of the solutions, focusing on two aspects.

\noindent{\textbf{[Time]}} To assess the ability of the solution to scale to large consumption series, 
we measured and compared the training time. 
In addition, as the Transformer architecture is known to have time complexity depending on the input sequence length, we also evaluate the gain of using slicing subsequences against the entire consumption series on the running inference performance.

\noindent{\textbf{[Memory]}} We measured the memory consumption of all the baselines inside and outside ADF during training. 
We tracked the total CPU memory for non-deep learning methods (Arsenal and ROCKET). 
For deep learning-based methods (that run on a GPU), we tracked the consumed GPU memory for a minibatch of data for a full forward/backward pass and an optimization step. 
We set the size of the minibatch to 32 for each baseline. 

\subsection{Results}

Table~\ref{table:CERresults} reports the results for the appliance detection cases for the CER dataset, while Table~\ref{table:EDFResults} reports the results obtained for the EDF 1 dataset.
For methods that use subsequences (Deng et al.~\cite{dengetal} and ADF combined with different classifiers), we report the scores for the parameter $w$ that leads to the best results (averaged over all cases).
First, note that ADF, when combined with various classifiers, outperforms those used outside of ADF, and outperforms the framework proposed by Deng et al.
Specifically, we observe an average increase for all the classifiers (Arsenal, ConvNet, ResNet, and Inception) when combining them with ADF of two and five points on the CER and EDF 1 datasets.
On average, ADF combined with the pretrained TransApp is better than all other classifiers on the two datasets (avg. score and avg. rank).
Furthermore, the larger TransApp (TransAppPT-l) pretrained on the (large) non-labeled EDF 2 dataset outperformed all other classifiers.
These results demonstrate that pretraining is an important step to significantly boost the model's performance.

\begin{figure}
    \centering
    \includegraphics[width=1.\linewidth]{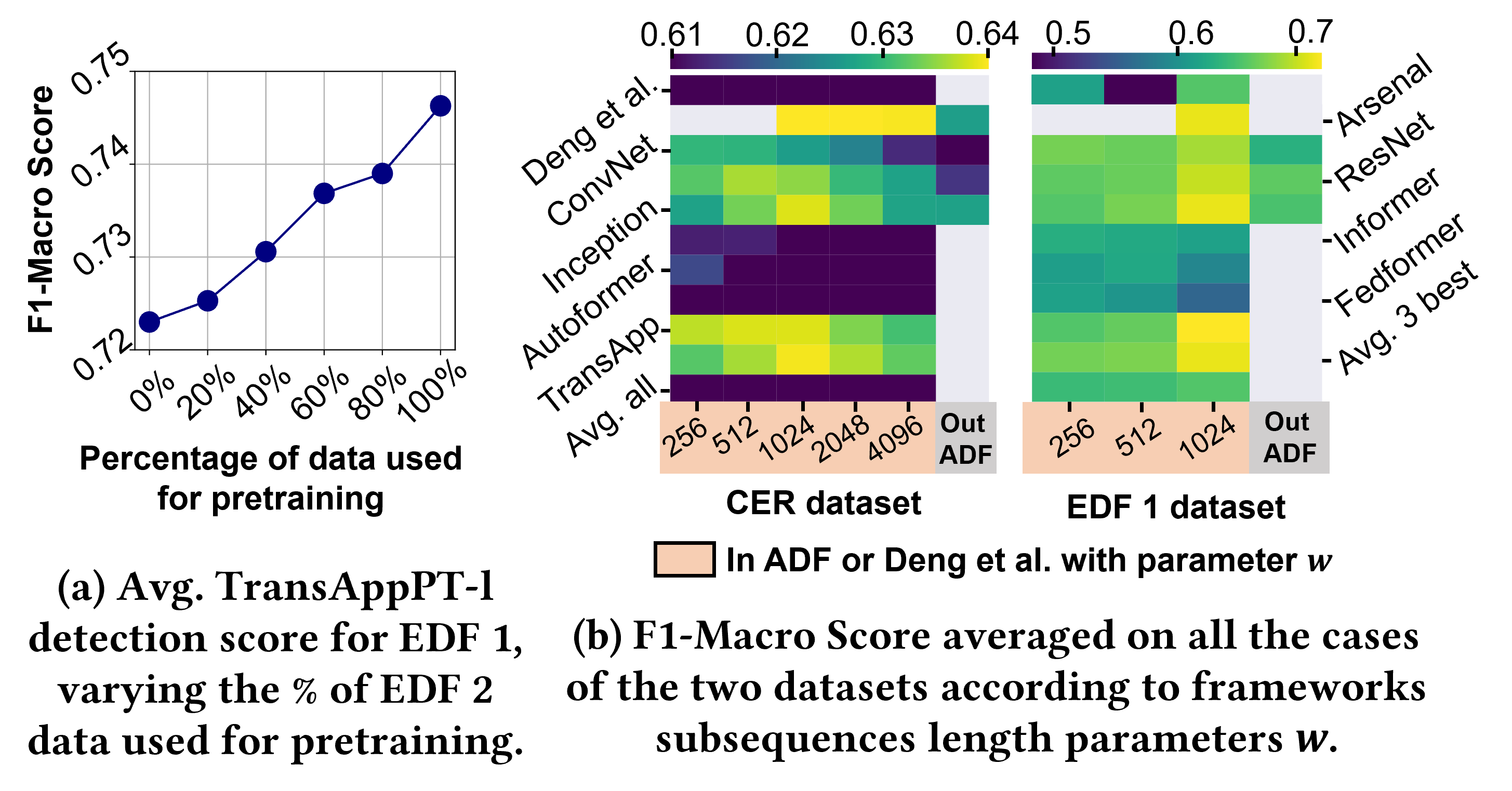}
    \vspace*{-0.7cm}
    \caption{(a) Impact of amount of unlabeled data used for pretraining; (b) Sensitivity study for $w$.}
    \label{fig:PretrainingImpactWinImpactCases}
\end{figure}

The results also imply that our solution can be used by suppliers in real scenarios to detect appliances.
Indeed, most appliances are accurately detected, with a Macro F1-Score higher than 0.8 for Electric Vehicles, Electric Heaters, and Water Heater appliances.
Figure~\ref{fig:PretrainingImpactWinImpactCases}(a) shows the average detection score for TransAppPT-l across all the cases of EDF 1, when varying the size of EDF 2 data used for pretraining.
More precisely, we pretrained with a random percentage of selected consumption series and finetuned it on all the detection cases (0\% means no pretraining).
The results show that the detection score increases proportionally to the amount of data used, confirming the validity of our proposed pretraining, and of using non-labeled data collected by electricity suppliers.

\noindent{\textbf{[Sensitivity influence of $w$]}} Figure~\ref{fig:PretrainingImpactWinImpactCases}(b) report the average sensitivity influence of the parameter $w$ over all the baselines for the two datasets.
Deng et al. and ADF combined with Informer, Autoformer, and Fedformer seem to benefit from small subsequences.
However, they achieve poor results compared to all the baselines regarding this parameter.
ConvNet and ResNet benefit from $w=512$ on the CER dataset.
However, all the 3 best competitors (Arsenal, Inception, and TransApp) benefit from $w=1024$ on the two datasets.

\begin{figure}
    \centering
    \hspace*{-0,2cm}
    \includegraphics[width=1.05\linewidth]{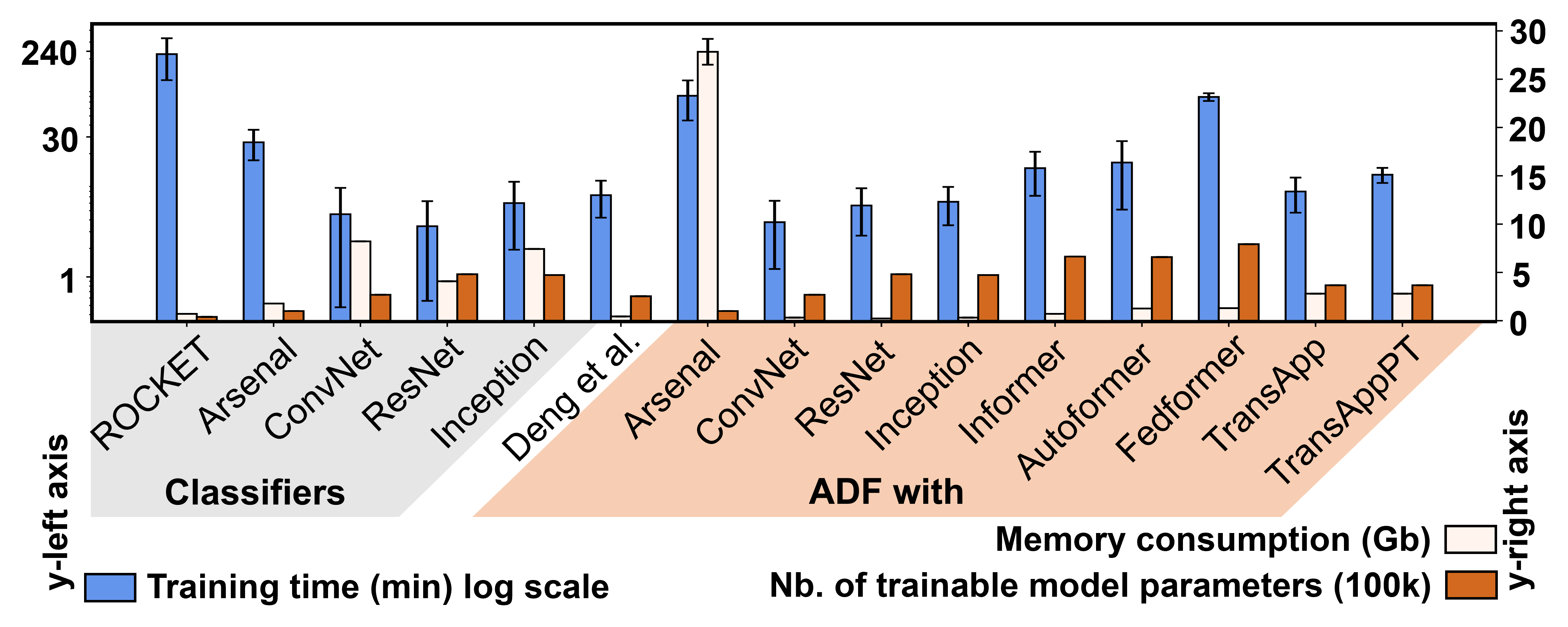}
    \vspace*{-0.6cm}
    \caption{Training time, memory consumption, and number of trainable parameters averaged across the different appliance detection cases for the CER dataset (log scale left y-axis).}
    \label{fig:trainingmemoryparamtime} 
    \vspace*{-0.2cm}
\end{figure}

\noindent{\textbf{[Training scalability Performance]}} Figure~\ref{fig:trainingmemoryparamtime} shows the training time cost, memory consumption, and number of trainable parameters of the different approaches (averaged across all possible detection cases of the CER dataset).
We show that for all the classifiers (except Arsenal), their training time is equivalent inside and outside ADF.
Moreover, we show that TransApp is equivalent to train as convolutional-based methods in ADF, and faster than other Transformer-based approaches.
Note that the reported training time for TransAppPT corresponds to the addition of the pretraining and finetuning steps.
Although Arsenal is the second best accurate on the CER dataset, this classifier's training time is very slow compared to other methods, especially when used inside ADF.
The training time of this classifier depends mainly on the number of instances rather than their entire length.
In addition, we note that except for Arsenal, all the baselines use less memory consumption when used inside ADF.
We also note that the memory consumption of TransApp, as well as its number of parameters, is kept small thanks to the use of ADF.

\begin{figure}
    \centering
    \includegraphics[width=0.8\linewidth]{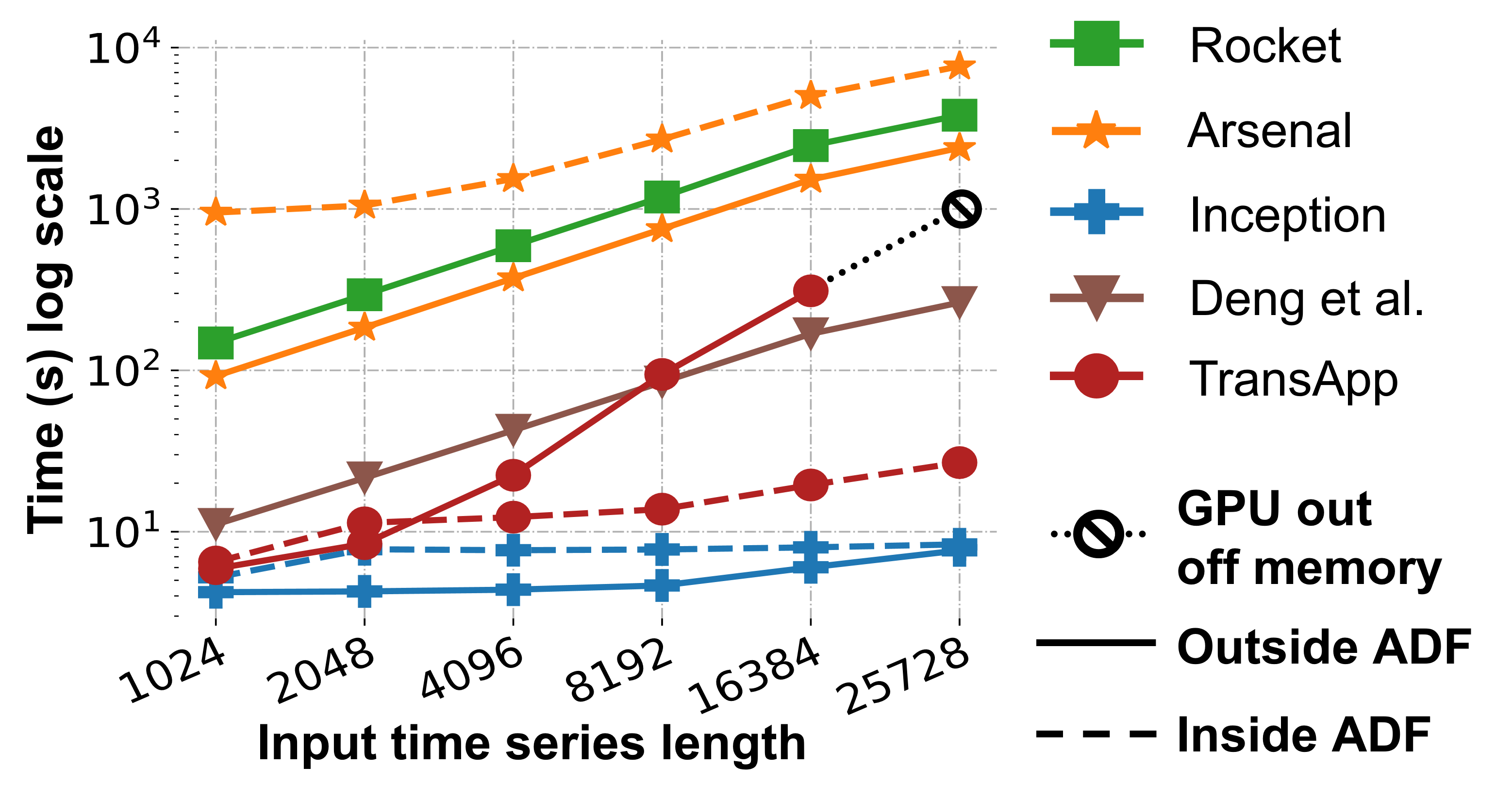}
    \vspace*{-0.2cm}
    \caption{\label{fig:inferencetime} Running time for classifiers (inside/outside ADF) to predict 1K instances labels according to the length of the input consumption series using the CER dataset.}
    \vspace*{-0.2cm}
\end{figure}

Figure~\ref{fig:inferencetime} shows the inference time to predict 1K consumption series labels according to the length of the input sequence.
A dashed line corresponds to using a specified classifier inside our proposed ADF. 
In contrast, a solid line corresponds to the same classifier that uses the entire series as inputs.
First, we can see that Inception (inside and outside our framework) is the fastest and is insensitive to the length of the whole input time series.
Secondly, we see that Arsenal and Rocket (the two random kernel convolution-based methods) are way slower at inference than other proposed approaches.
However, our proposed TransApp model inside our framework can also scale nearly linearly according to the sequence length.
However, we show that using our TransApp classifier outside our framework makes it sensitive to the entire input length and, therefore, leads to a much longer inference time for long sequences.

Overall, we show that the training and running time of our proposed ADF approach combined with our TransApp classifier can scale as efficiently as other approaches using long sequences and large databases currently available by electricity suppliers.

\subsection{Ablation Study and Hyperparameters}
\label{sec:AblationStudy} 


\noindent{\textbf{[Ablation Studies]}} Table~\ref{table:AblationStudy} shows the average F1-Macro Scores for the CER and EDF 1 datasets for the non-pretrained TransApp inside ADF when we remove key elements of the proposed architecture.
"Default" shows the score obtained with the base architecture (i.e., the same results as the one obtained in Table~\ref{table:CERresults} and Table~\ref{table:EDFResults}). 
Then, we show the results with the three following parts removed successively:
(i) Embedding block: we replace the Convolutional Embedding Block with a simple linear embedding layer that maps each time step of the input sequence $\boldsymbol{x}^{w \times m}$ to a space $\boldsymbol{x}^{m \times d_{\text{model}}}$ (the approach used in~\cite{zerveas_tst}).
(ii) DMSA: we replace the Multi-Head DMSA with the original Multi-Head Self-Attention mechanism.  
(iii) Time Encoded features: We use only the univariate consumption time series as input for training our TransApp architecture.

The results show that the Embedding Block is the most important part of our proposed architecture and therefore confirms the need to introduce an inductive bias to render the Transformer able to detect the different appliance patterns more accurately.
Subsequently, we see that removing the DMSA mechanism or not using temporal encoded features leads to a small drop in the average results.

\noindent{\textbf{[Hyperparameter Choice]}} Tables~\ref{table:hyperparameters}(a) and (b) show the results of the choice of two important parameters.
Table~\ref{table:hyperparameters}(a) shows that the best inner dimension $d_{\text{model}}$ is 96, while 
Table~\ref{table:hyperparameters}(b) motivates the choice of not using positional encoding before the Transformer block (cf. Section~\ref{subsec:transapparchitecture}), as is usually done in Transformer architectures.
The findings indicate that utilizing a fixed positional encoding (as originally proposed~\cite{attentionisallyouneed}), 
or a fully learnable one (as proposed in~\cite{zerveas_tst}) results in a decrease in the model's average performance.

\noindent{\textbf{[Effect of the pretraining on the baselines]}} We performed experiments to evaluate the effect of our proposed pretraining process on other deep-learning competitors (ConvNet, ResNet, and Inception).
However, these baselines do not benefit from our two-step training, 
as the scores reached by the pretrained version of the model were equivalent to the ones obtained with the non-pretrained.

\begin{table}
\caption{\label{table:AblationStudy} Average F1-Macro Score results for the CER and EDF 1 datasets when removing key parts of ADF \& TransApp.}
\vspace{-0.1cm}
{\footnotesize
\begin{tabular}{c|ll}
\toprule
    \textbf{TransApp} & \textbf{CER} & \textbf{EDF 1} \\
    \hline
    Default & \textbf{0.633} & \textbf{0.723} \\
   \textbf{w/o} Embedding Block & 0.594 (\textit{-6.2 \%}) & 0.659 (\textit{-8.9 \%})\\
   \textbf{w/o} DMSA & 0.623 (\textit{-1.6 \%}) & 0.717 (\textit{-0.8 \%})\\
   \textbf{w/o} Time Encoded Features & 0.625 (\textit{-1.3 \%}) & 0.719 (\textit{-0.6 \%})\\
\bottomrule
\end{tabular}
}
\end{table}

\begin{table}
\caption{Influence of hyperparameters choice on the average detection score on the CER and EDF 1 datasets.}
\vspace{-0.2cm}
\label{table:hyperparameters} 
\centering
{\footnotesize
\subfloat[Inner dimension]{%
\begin{tabular}{c|cc}
\toprule
\textbf{Dim} & \textbf{CER} & \textbf{EDF 1} \\
\hline
64 & 0.62 & 0.719 \\
\textbf{96} & \textbf{0.633} & \textbf{0.723} \\
128 & 0.62 & 0.717 \\
\bottomrule
\end{tabular}}%
\qquad\qquad
\subfloat[Positional Encoding]{%
\begin{tabular}{c|cc}
\toprule
\textbf{Type} & \textbf{CER} & \textbf{EDF 1} \\
\hline
\textbf{None} & \textbf{0.633} & \textbf{0.723} \\
Fixed & 0.623 & 0.711 \\
Learnable & 0.620 & 0.715 \\
\bottomrule
\end{tabular}}%
}
\vspace{-0.1cm}
\end{table}

\section{Conclusions}
\label{sec:concl}

This paper proposes the ADF framework and the TransApp classifier to detect appliances in very low-frequency electricity consumption time series.
The classifier is first pretrained in a self-supervised manner to enhance its performance on appliance detection tasks.
The results show that ADF improves the performance of state-of-the-art time series classifiers applied to appliance detection and that the TransApp architecture is the most accurate.
Furthermore, we show that ADF enables TransApp to scale to very long consumption series, making it applicable to real-world scenarios. 


\bibliographystyle{ACM-Reference-Format}
\bibliography{sample}

\end{document}